\documentclass{mnras}
\usepackage{graphicx}

\def\simlt{\ {\raise-.5ex\hbox{$\buildrel<\over\sim$}}\ }

\def\simgt{\ {\raise-.5ex\hbox{$\buildrel>\over\sim$}}\ }

\def\cd{d$^{-1}$\,}

\begin{document}

\title[BRITE and ground-based photometry of $\nu$~Eridani]
{Combining BRITE and ground-based photometry for the $\beta$~Cephei star 
$\nu$~Eridani: impact on photometric pulsation mode identification and 
detection of several g modes\thanks{Based on data collected by the BRITE 
Constellation satellite mission, designed, built, launched, operated and 
supported by the Austrian Research Promotion Agency (FFG), the University 
of Vienna, the Technical University of Graz, the Canadian Space Agency 
(CSA), the University of Toronto Institute for Aerospace Studies (UTIAS), 
the Foundation for Polish Science \& Technology (FNiTP MNiSW), and 
National Science Centre (NCN).}}
\author[G. Handler et al.]
{G. Handler,$^{1}$ M. Rybicka,$^{1}$ A. Popowicz,$^{2}$ A. Pigulski,$^{3}$ 
R. Kuschnig,$^{4,5}$ \and E. Zoc\l{}o\'nska,$^{1}$ A. F. J. 
Moffat,$^{6}$ W. W. Weiss,$^{4}$ C. C. Grant,$^{7}$ H. Pablo,$^{6}$ \and 
G. N. Whittaker,$^{8}$ S. M. Ruci\'nski,$^{8}$ T. Ramiaramanantsoa,$^{6}$
K. Zwintz,$^{9}$ G. A. Wade\,$^{10}$
 \and \\
$^1$ Nicolaus Copernicus Astronomical Center, Bartycka 18, 00-716 Warsaw, 
Poland\\
$^{2}$ Silesian University of Technology, Institute of Automatic Control, 
Gliwice, Akademicka 16, Poland\\
$^{3}$ Instytut Astronomiczny, Uniwersytet Wroc\l{}awski, ul. Kopernika 11,
51-622 Wroc\l{}aw, Poland\\
$^4$ Institute for Astrophysics, Universit\"at Wien, T\"urkenschanzstrasse
17, A-1180 Wien, Austria\\
$^5$ Institut f\"ur Kommunikationsnetze und Satellitenkommunikation, 
Inffeldgasse 12/I, 8010 Graz, Austria\\
$^{6}$ D\'epartment de physique, Universit\'e de Montr\'eal, C. P. 6128, 
Succ. Centre-Ville, Montr\'eal, QC H3C 3J7, Canada\\
$^{7}$ Space Flight Laboratory, University of Toronto, 4925 Dufferin 
Street, Toronto, M3H 5T6, Canada\\
$^{8}$ Department of Astronomy and Astrophysics, University of
Toronto, 50 St. George Street, Toronto, ON M5S 3H4, Canada\\
$^{9}$ Institut f\"ur Astro- und Teilchenphysik, Universit\"at Innsbruck, 
Technikerstrasse 25/8, 6020, Innsbruck, Austria\\
$^{10}$ Department of Physics, Royal Military College of Canada, PO Box 
17000, Stn Forces, Kingston, ON K7K 7B4, Canada}

\date{Accepted 2016 July 17.
  Received 2016 August 13;
  in original form 2016 September 10}
\maketitle
\begin{abstract} We report a simultaneous ground and space-based 
photometric study of the $\beta$~Cephei star $\nu$~Eridani. Half a year 
of observations have been obtained by four of the five satellites 
constituting BRITE-Constellation, supplemented with ground-based 
photoelectric photometry. We show that carefully combining the two data 
sets virtually eliminates the aliasing problem that often hampers 
time-series analyses. We detect 40 periodic signals intrinsic to the star
in the light curves. Despite a lower detection limit we do not recover 
all the pressure and mixed modes previously reported in the literature, 
but we newly detect six additional gravity modes. This behaviour is a 
consequence of temporal changes in the pulsation amplitudes that we also 
detected for some of the p modes. We point out that the dependence of 
theoretically predicted pulsation amplitude on wavelength is steeper in 
visual passbands than those observationally measured, to the extent that 
the three dominant pulsation modes of $\nu$ Eridani would be incorrectly 
identified using data in optical filters only. We discuss possible 
reasons for this discrepancy.
\end{abstract}

\begin{keywords}
stars: variables: other -- stars: early-type -- stars: oscillations
-- stars: individual: $\nu$~Eridani -- techniques: photometric
\end{keywords}

{\section{Introduction}}

$\nu$~Eridani (HD 29248, $V=3.92$, B2 III, henceforth $\nu$~Eri) is 
arguably the asteroseismically best studied $\beta$~Cephei star (for 
more information on this group of massive pulsating variable stars see 
Stankov \& Handler 2005). After an immense ground-based observational 
effort involving both multisite photometry and high-resolution 
spectroscopy (Handler et al. 2004, Aerts et al. 2004, De Ridder et al. 
2004, Jerzykiewicz et al. 2005, hereinafter JHS), tight constraints on 
the overall chemical composition and convective core overshooting of 
this star were obtained, in combination with a detection of differential 
internal rotation (Pamyatnykh, Handler \& Dziembowski 2004). The latter 
study as well as alternative models to explain the star's pulsation 
spectrum (Ausseloos et al. 2004) additionally pointed towards a 
necessity for an increase of the iron-peak element opacities.

This success was possible because the star's pulsation spectrum happened 
to be very suitable for asteroseismic studies. The pulsations contain a 
dominant radial mode that immediately constrains the stellar mean 
density. Furthermore, a triplet of dipole modes was identified 
observationally, two more $l=1$ multiplets were found, and a single 
quadrupole mode was detected. Fitting this observed pulsation spectrum 
with theoretical models (Pamyatnykh et al. 2004, Ausseloos et al. 2004) 
revealed that the two lowest-frequency $l=1$ multiplets correspond to 
mixed modes, with different contributions from the gravity and pressure 
mode cavities, which makes the seismic information from them largely 
independent and complementary. 

In addition, low-frequency oscillations likely due to stellar gravity 
modes were detected (Handler et al. 2004, Aerts et al. 2004, JHS). 
However, finding excitation of the latter in corresponding stellar 
models proved to be impossible and would require additional modification 
to the opacities used for the stellar models (Dziembowski \& Pamyatnykh 
2008). Finally, Daszy{\'n}ska-Daszkiewicz \& Walczak (2010) applied 
their method of complex asteroseismology to $\nu$ Eri and showed 
that neither with OPAL nor with OP opacities alone the observed complex 
amplitude of the bolometric flux variation can be reproduced for both 
its p and g modes at the same time.

After these studies, the asteroseismic information on the star appeared 
to be made full use of, at least as far as studies hinged on ground-based 
data are concerned: the above-mentioned campaigns yielded some 1130 h of 
time-resolved photometry and 430~h of time-series spectroscopy. 
Therefore, a return to $\nu$~Eri only appeared sensible once new data 
sets, extensive enough to resolve the pulsation spectrum, and of a 
quality allowing to decrease the noise level significantly would become 
accessible.

Such data sets are available nowadays. BRITE-Constellation (Weiss et al. 
2014) is a set of five nearly identical nanosatellites with high-quality 
pointing stability, each hosting a telescope with 30~mm aperture and an 
uncooled 11 Megapixel CCD, which allows one to acquire time-series 
photometry of $V \simlt 4.5$ targets in a 24 degree field of view. Two 
broadband filters are available, a blue filter with a central wavelength 
of about 420 nm and a red filter centred at 620 nm. Although these 
filters do not have the standard Johnson-Cousins bandpasses, their 
effective wavelengths are similar to those of Johnson $B$ and Cousins 
$R_c$, and we therefore name them simply $B$ and $R$ in the remainder of 
this work.

The satellites are in low-Earth orbits, meaning that they can observe 
the target fields for up to six months a year and for about 20 minutes 
per 101-minute orbit, depending on their position with respect to the 
Sun and the Earth. $\nu$~Eri was observed during the second 
BRITE-Constellation campaign directed towards the field of Orion. In 
anticipation of these measurements, ground-based photometric 
observations were organized to coincide temporally. In the following, we 
report the results of both ground and space-based monitoring.

\vspace{4mm}

\section{Observations and reductions}

\subsection{Ground-based photometry}

Photoelectric time-series measurements in support of BRITE were 
organized for two reasons. First, their analysis can be directly 
compared with that of JHS, and second, near-ultraviolet photometry that 
BRITE cannot provide, is invaluable for the identification of the 
pulsation modes of $\beta$~Cep stars. Our ground-based data were 
obtained at two observatories. The bulk of the measurements originated 
from the 0.75-m Automatic Photoelectric Telescope (APT) T6 at Fairborn 
Observatory in Arizona. In addition, the 0.5-m telescope at the South 
African Astronomical Observatory was employed (observer EZ). 
Differential time-series photoelectric data were collected through the 
Str\"omgren $uvy$ filters, and the ``classical'' comparison stars 
$\mu$~Eri (HD 30211, B5IV, $V=4.00$) and $\xi$~Eri (HD 27861, A2V, 
$V=5.17$) were used. Table~\ref{tab:glog} gives a summary of those 
observations; the time base is 135~d.

\begin{table}
\caption[]{Log of the ground-based photometry of $\nu$~Eri.}
\begin{center}
\begin{tabular}{lcccccl}
\hline
Observatory & $T_{\rm start}$ & $T_{\rm end}$ & \# data\\
& \multicolumn{2}{c}{HJD -- 2456000} \\
\hline
Fairborn & 943.800 & 1078.710 & 1429\\
SAAO & 996.319 & 1019.347 & 131\\
\hline
\end{tabular}
\end{center}
\label{tab:glog}
\end{table}

The data were reduced in a standard way, starting by correcting for 
coincidence losses, sky background and extinction. Standard extinction 
coefficients for the observing sites were used. As $\mu$~Eri is a known 
variable star (e.g., see Jerzykiewicz et al. 2013), and has also been 
observed simultaneously by BRITE-Constellation, these data were set 
aside for a separate analysis and only the differential magnitudes 
($\nu$~Eri $-$ $\xi$~Eri) were analysed further. As a final step, the 
timings for this differential light curve were corrected to the 
heliocentric frame.

Given that there was no constant star in the observing sequence, we 
cannot directly specify the accuracy of these observations. However, 
experience with measurements taken under similar conditions (e.g., 
Handler et al. 2012) suggests that the rms accuracy per single data 
point should be well below 4 ($u$ filter) and 3 mmag ($v$ and $y$ 
filter), respectively.

\subsection{BRITE-Constellation}

The data were obtained with four out of the five satellites: 
BRITE-Austria (hereinafter BAb), BRITE-Lem (BLb), BRITE-Heweliusz (BHr), 
and BRITE-Toronto (BTr). The first two satellites observe through the 
blue filter, the latter two use the red one (hence the suffixes ``b'' 
and ``r'' in their respective abbreviations). The data for each target 
are downloaded from the satellites in subrasters of (usually) $28 \times 
28$ pixel size. Initial photometric reductions are carried out with a 
pipeline that takes into account bad pixels, median column counts, image 
motion and PSF variations (see Pablo et al. 2016, Popowicz 2016 and 
Popowicz et al. 2016).

The data reported here were delivered to the user in the form of ASCII 
files that contain the HJD of the observations, stellar flux counts, x 
and y position of the stellar profile on the raster, CCD temperature and 
a quality flag. The information aside from HJD and counts is required 
for data decorrelation. In brief, pixel-to-pixel and intrapixel 
sensitivity variations modify the measured stellar flux depending on the 
x/y position on the chip. Temperature variations within the satellites 
and hence also of the detector during the orbit and on longer time 
scales also introduce apparent modulations in the signal derived from 
the initial CCD photometry. These effects may also be intermingled, as 
the temperature changes also modify the optical system, and the stellar 
PSFs are large and of varying shape. Therefore, temperature-induced 
changes of the PSF also modify the relation between the x/y position and 
the distribution of the PSF over the individual pixels (discussed in 
detail by Buysschaert et al. 2016). Therefore, BRITE photometry needs to 
be carefully decorrelated for these factors before proceeding to 
scientific analyses of the intrinsic stellar variability. The basic 
process of decorrelation of BRITE data has been described in detail by 
Pigulski et al. (2016).

In case of constant stars or short-period low-amplitude variables, 
decorrelation can be performed directly on the data. However, for 
$\nu$~Eri intrinsic stellar variability dominates the light curves. 
Therefore, decorrelation was carried out in two steps. First, bad points 
were discarded (according to the data quality flags) and a preliminary 
frequency analysis was carried out on the light curves from the 
individual satellites. A multifrequency fit consisting of the detected 
signals was determined and subtracted from the data. Consequently, the 
residuals were inspected for correlations, and the latter removed on a 
step-by-step basis. We have largely followed the procedures laid out by 
Pigulski et al. (2016), with a few deviations. First, $\nu$~Eri is not 
bright enough to require a correction for CCD nonlinearity. Second, 
whenever we noticed a temperature-dependent change of the correlations 
with x/y position we separated the data into smaller temporal subsets 
that were decorrelated individually. Third, in two cases an additional 
correction of the light curve was necessary due to variable background 
illumination due to scattered light, e.g. from the Moon. Finally, a few 
significant outliers were rejected. The rejected points had a 
probability of less than 50\% to fall within the standard deviation of 
all data points. These outliers are suspected to be caused by blurred 
images or strong cosmic ray hits, and are very small in number ($<$ 0.2\% 
of all data points).

An overview of the characteristics of the data sets used for analysis is 
given in Table~\ref{tab:blog}; the time base of the BRITE observations 
is somewhat above 170~d in both filters (including a $\approx 45$\,d gap 
in the blue-filter data). The precision of the measurements as given in 
this table was deduced from the rms scatter of the orbital means of the 
pulsation-filtered decorrelated light curves and must therefore be 
considered a lower limit to the real accuracy. Even though the 
measurements by the different satellites have varied quality, we decided 
against the application of statistical weights to the different data 
sets.

Due to the complexity of this reduction procedure, special care was 
taken to test its effects on the frequency analysis to follow. In other 
words, we checked whether the decorrelation could introduce spurious 
periodic signals. The first test consisted of introducing synthetic 
periodic signals into the BRITE light curves of $\nu$~Eri at frequencies 
deemed to be critical (the possible g-mode frequency domain between 0.1 
- 0.9\,\cd, the Earth's rotation frequency 1\,\cd, and frequencies close 
to the known stellar oscillations). These modified light curves were 
decorrelated like the original ones. The result of this experiment was 
that the amplitudes of some of the artificial signals were suppressed 
(mostly within the errors), and that no additional spurious signals 
originated. The second test was to decorrelate the BRITE data of another 
star observed in the same field, $\gamma$~Ori, deemed to be constant, 
the same way as the $\nu$~Eri light curves. The aim of this test was to 
check which frequency domains can be affected by imperfections in the 
data decorrelation process. Even though the exact correlations in the 
data may not necessarily be the same due to the variation of the PSF 
over the BRITE field of view (Pablo et al. 2016), the time scales at 
which reduction deficiencies will manifest themselves should be the 
same. We will compare the results for $\gamma$~Ori and $\nu$~Eri in the 
next section.

\begin{figure*}
\includegraphics[width=181mm,viewport=-35 15 570 350]{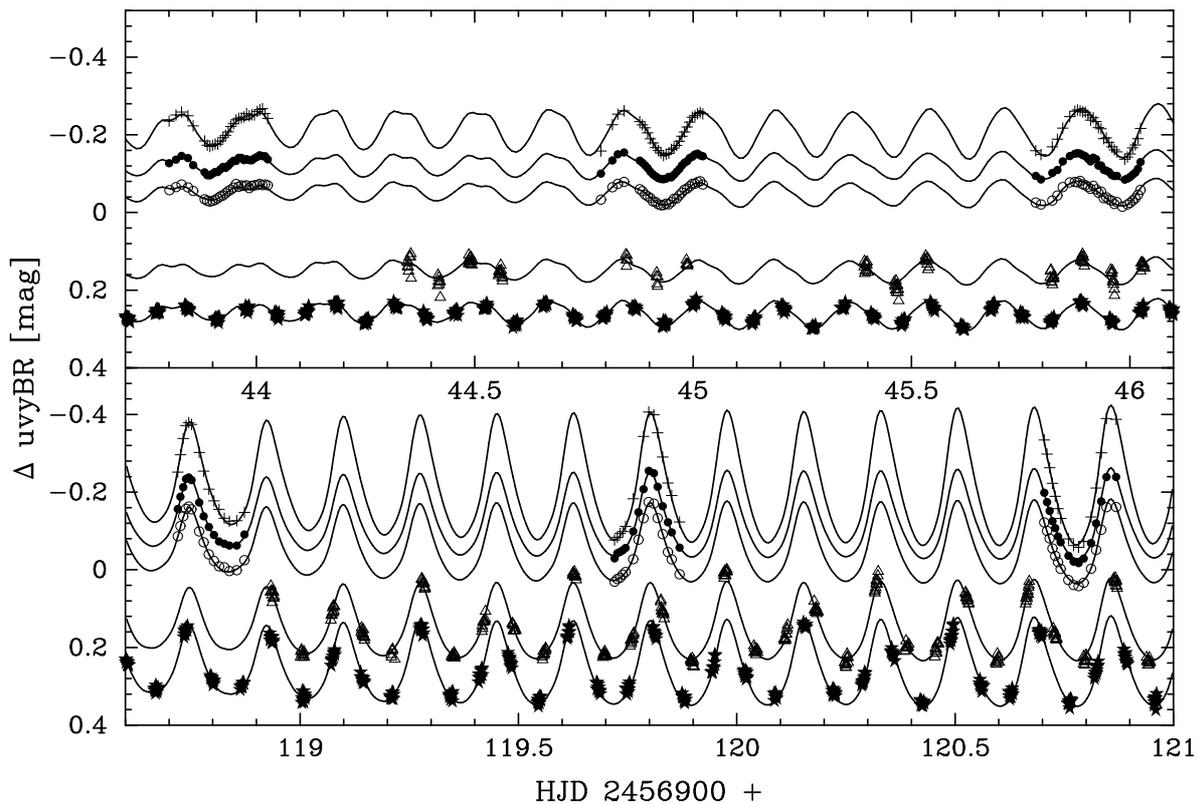}
\caption[]{Some example light curves of $\nu$ Eri from our observations. 
The plus signs represent the $u$ filter data, the full circles the $v$ 
filter light curves, and the open circles the $y$ filter light curves. 
The BRITE $B$ observations are the open triangles (1-minute averages, 
upper panel: BAb, lower panel BLb), and the BRITE $R$ observations 
marked with star symbols (1-minute averages, upper panel: BTr, lower 
panel BHr). The lines are multifrequency fits to the light curves to be 
computed in Sect. 3.}
\label{fig:lcs}
\end{figure*}

Finally, part of the BAb measurements were of considerably poorer 
quality than the remainder of the BRITE photometry (above 8 mmag rms in 
the orbital means) due to temporal pointing instabilities. Hence these 
observations were not analysed further. For frequency analysis, all data 
were binned into one-minute and orbital means (if a sufficient number of 
observations was obtained during any given orbit). Some example light 
curves are shown in Fig.~\ref{fig:lcs}, that also demonstrates the 
different sampling of the space and ground-based data and the changes in 
the amplitude due to the beating of multiple pulsation periods.


\begin{table}
  \caption[]{Log of the BRITE measurements of $\nu$~Eri. Start and end times of the individual satellites' observations are given as well as the number of one-minute data bins, the number of orbits observed, and the precision of the orbital means.}
\begin{center}
\begin{tabular}{lcccccl}
\hline
Satellite & $T_{\rm start}$ & $T_{\rm end}$ & \multicolumn{2}{c}{\# data} & 
$\sigma$\\
& \multicolumn{2}{c}{HJD -- 2456000} & 1-min & orb & mmag\\
\hline
BAb & 926.354 & 953.075 & 1453 & 198 & 4.6\\
BLb & 998.527 & 1098.276 & 12536 & 1089 & 1.9\\
BTr & 924.724 & 995.525 & 11753 & 676 & 1.1\\
BHr & 972.718 & 1095.509 & 12802 & 1107 & 2.3\\
\hline
\end{tabular}
\end{center}
\label{tab:blog}
\end{table}

\section{Frequency analysis}

We analysed our data using the {\sc Period04} software (Lenz \& Breger 
2005). This package applies single-frequency power spectrum analysis and 
simultaneous multi-frequency sine-wave least-square fitting. It also 
includes advanced options such as the calculation of optimal light-curve 
fits for multiperiodic signals including harmonic, combination, and 
equally spaced frequencies.

To assess the detection significance of any given signal, we adopted the 
widely used and reliable criterion by Breger et al.\ (1993, 1999). 
According to this criterion, any independent peak that exceeds a 
signal-to-noise (S/N) ratio of four in the amplitude spectrum is 
statistically significant, whereas any combination (sum or difference) 
frequency is considered detected when exceeding $S/N=3.5$. To compute 
the noise level in the presence of frequency-dependent red noise, we 
integrated the amplitude spectrum in sliding windows over intervals of 
2\,\cd.

\begin{figure}
\includegraphics[width=84mm,viewport=00 05 270 511]{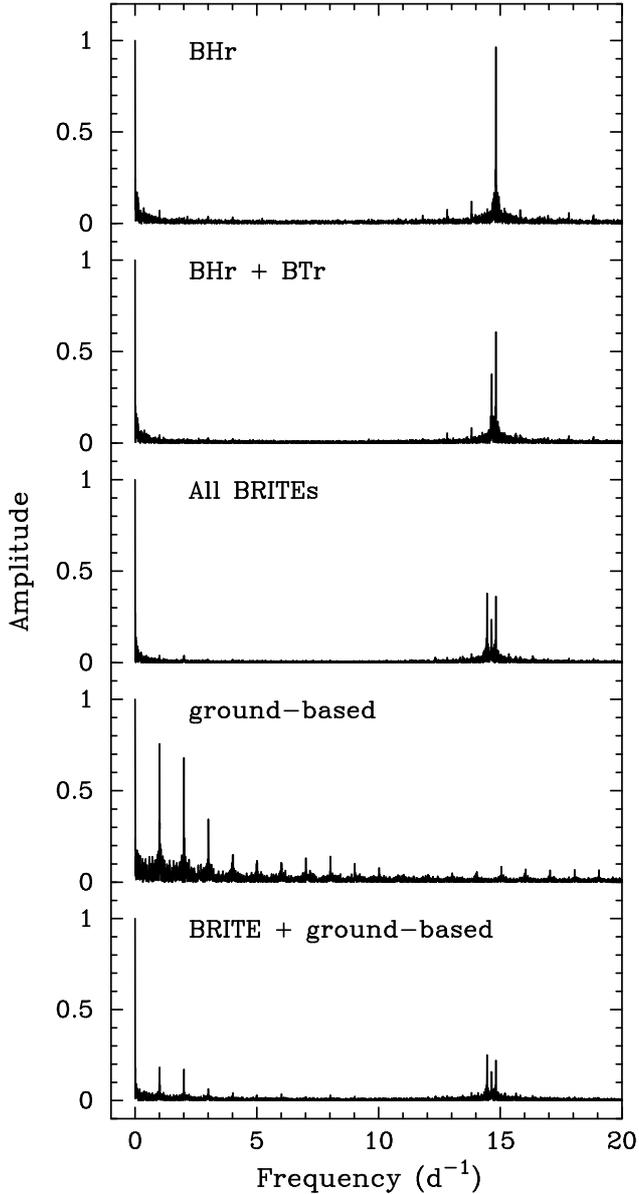}
\caption[]{Uppermost panel: spectral window function of the BHr data only. 
Second panel: the same, but for the combined red-filter BRITE data. Third 
panel: the same, with all red- and blue-filter BRITE data combined. Fourth 
panel: spectral window for the ground-based $y$-filter measurements. Fifth 
panel: spectral window for the merged BRITE and ground-based data.}
\label{fig:spws}
\end{figure}

However, before actually looking for pulsational signals in the light 
curves, it is worthwhile to consider the characteristics of the data at 
hand. The BRITE observations are very close to being evenly spaced, 
causing strong aliasing at the orbital frequency. An example is shown in 
the upper panel of Fig.~\ref{fig:spws}, which contains the spectral 
window function of the measurements by BHr. The alias peak at the 
orbital frequency has 96\% of the amplitude of the zero-frequency input 
signal. Therefore analysing the data from a single satellite can easily 
result in mistakes in frequency detection.

This problem can be mitigated when data from more than one satellite are 
available. In the second panel of Fig.~\ref{fig:spws} the spectral window 
function of the combined data from the two red-filter BRITEs is shown. The 
alias at the orbital frequency is clearly diminished, as a consequence of 
the different orbits of the two satellites: their orbital periods are not 
exactly the same, and they do not observe at the same time. 
Figure~\ref{fig:lcs} also illustrates this fact: in the lower panel it can 
be easily discerned that the observation window of BLb drifts with respect 
to that of BHr. Returning to Fourier space (Fig.~\ref{fig:spws}, third 
panel), the orbital aliases can be suppressed even more when the data from 
all four BRITEs employed to observe $\nu$~Eri are combined.

As it is well known, a similar problem is present in ground-based 
observations due to the regular daytime gaps in the measurements. In the 
fourth panel of Fig.~\ref{fig:spws} we show the spectral window function 
of our ground-based observations; the well-known daily aliasing pattern is 
clearly present. It should be noted that adding the comparably small data 
set from SAAO to the APT data resulted in a decrease of the 1\cd alias 
from 91\% to 76\% of the central peak.

Because the aliasing structure of the space observations differs strongly 
from that of ground-based data, combining them results in an even stronger 
depression of the aliases (lowest panel of Fig.~\ref{fig:spws}). 
Therefore, from the point of view of frequency detection it would be 
advantageous if such a combined data set could be exploited.

We thus examined how such a data set could be created, given that the 
amplitudes of the stellar variability are different from filter to 
filter, and that phase lags may occur between the light curves from 
different filters. Fortunately, for $\beta$~Cep stars, both those 
amplitude differences and phase lags are not strong in the visual band. 
After some trials, we found that for $\nu$ Eri combining the BRITE 
measurements from all satellites with the ground-based $y$-filter data - 
for the purpose of frequency detection only - is the most useful 
approach. We note that for other stars or data/filter combinations the 
situation may be different and such an approach must always be made with 
great caution.

Consequently, we adopted the following frequency analysis strategy: we 
computed amplitude spectra of the combined data (using orbital means of 
the BRITE measurements) and identified the significant signals step by 
step. These were fitted to the data, using the 1-minute binned BRITE data 
to not falsify the amplitudes of the higher-frequency signals. We adopted 
the same frequencies for all filters, but determined the amplitudes and 
phases from the data sets in the individual filters separately before 
computing orbital averages and combining the residuals again. Then we 
searched for more stellar variability frequencies; we call this procedure 
prewhitening. The parameters of the newly detected periodic signals were 
optimized together with those found previously. Once no significant 
variation was left in the residuals, the analysis was terminated.

\begin{figure*}
\includegraphics[width=181mm,viewport=-35 15 570 740]{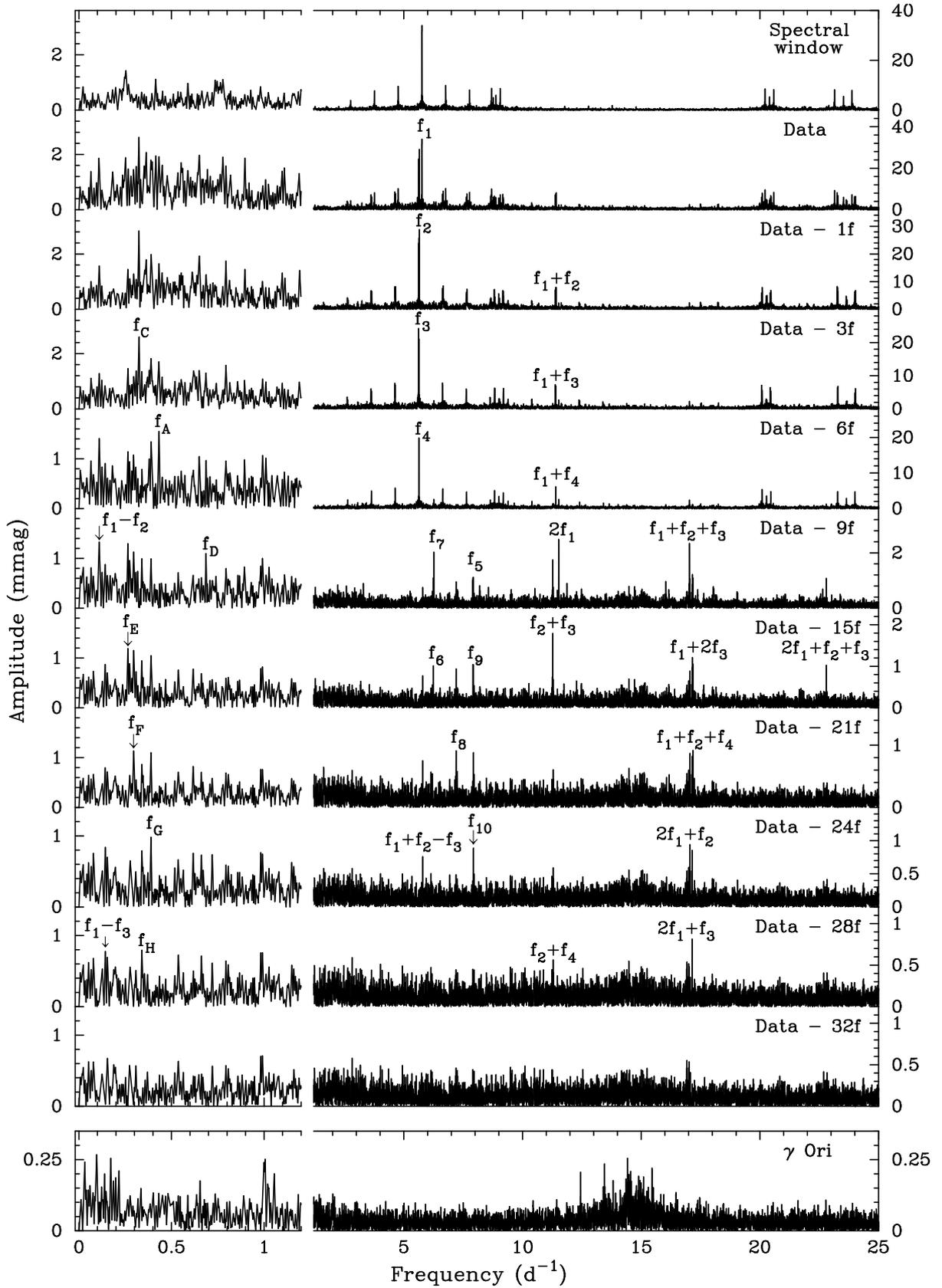}
\caption[]{Amplitude spectra of $\nu$~Eri. The uppermost panel shows 
the spectral window of the data, followed by the periodogram of the data. 
Successive prewhitening steps are shown in the following panels; note 
their different ordinate scales. Also note that the low-frequency domain 
is shown in higher resolution, and that the lowest panel shows the 
amplitude spectrum of the supposedly constant star $\gamma$~Ori for
comparison. The latter is not as ``flat'' as it is dominated by
reduction imperfections rather than by white noise.} 
\label{fig:prew}
\end{figure*}

Figure~\ref{fig:prew} illustrates this process. In the uppermost panel a 
spectral window is shown as the Fourier transform of a single noise-free 
sinusoid with a frequency of 5.7633 \cd (the strongest pulsational signal 
of $\nu$~Eri) and an amplitude of 36 mmag sampled in the same way as were 
our measurements. As expected, aliasing is strongly suppressed.

The amplitude spectra of the data itself (second panel of 
Fig.~\ref{fig:prew}) shows the signal designated $f_1$ as the strongest, 
but some additional structures not compatible with spectral window 
sidelobes are also present. Consequently, we prewhitened this signal as 
described before, and computed the amplitude spectrum of the residual 
light curve (third panel of Fig.~\ref{fig:prew}).

This resulted in the detection of a second signal ($f_2$) and of another 
variation at the sum frequency of the two previously detected. We then 
prewhitened a three-frequency fit from the data using the same 
optimisation method as before and fixed the combination term to the 
exact sum of the two independent frequencies within {\sc Period04}. The 
next panels show the continuation of this process until the detection of 
32 frequencies, all of which are significant within the criteria 
discussed before.

The lowest panel of Fig.~\ref{fig:prew} shows the amplitude spectrum of 
the BRITE data for $\gamma$~Ori, a constant star as far as we can tell. 
Comparing it with the amplitude spectrum of $\nu$~Eri prewhitened by 32 
frequencies, it can be concluded that possible artifacts left behind in 
the reduction procedure only seem to occur in confined frequency ranges: 
below 0.2\,\cd, between $0.95 - 1.1$\,\cd, and around the orbital 
aliases of these frequency domains (or vice versa). The frequencies 
detected for $\nu$~Eri are outside these domains, with the exception of 
the two lowest-frequency combinations. However, as these are also 
present in the ground-based data, there is no reason to doubt their 
reality.

The final results of the frequency analysis of our $\nu$~Eri data, with 
the amplitudes determined for the individual filters separately, are 
listed in Table~\ref{tab:freq}. This listing contains 40 signals, which 
is more than the 32 frequencies prewhitened in Fig.~\ref{fig:prew}. The 
additional eight signals are low-amplitude combination frequencies only. 
The designation of the independent frequencies was chosen to be 
consistent with JHS and Handler et al. (2004).

\begin{table*}
\caption[]{Multifrequency solution for our time-resolved photometry of
$\nu$~Eri. Formal error estimates (following Montgomery \& O'Donoghue
1999) for the independent frequencies are also quoted. Formal errors on 
the amplitudes are $\pm$ 0.3 mmag in $u$ and $\pm$ 0.2 mmag in $v$, 
$y$, $B$, and $R$. These formal errors are known to underestimate the real 
errors by about a factor of two (Handler et al. 2000, JHS). The S/N ratio 
quoted is for the data set used to create Fig.~\ref{fig:prew}. The 
detected signals are separated in groups of independent pressure and mixed 
modes (numerical subscripts), gravity modes (letter subscripts), as well
as first, second and third-order combination frequencies, respectively.} 
\begin{center}
\begin{tabular}{lccccccc}
\hline
ID & Freq. & $u$ Ampl. & $v$ Ampl. & $y$ Ampl. & $B$ Ampl. & $R$ Ampl. & $S/N$ \\
 & (\cd) & (mmag) & (mmag) & (mmag) & (mmag) & (mmag) \\
\hline
$f_{1}$ & 5.763264 $\pm$ 0.000012 & 68.0 & 38.3 & 34.4 & 37.6 & 33.0 & 262.8 \\ 
$f_{2}$ & 5.653917 $\pm$ 0.000015 & 39.6 & 27.5 & 26.2 & 27.7 & 25.8 & 195.9 \\ 
$f_{3}$ & 5.620047 $\pm$ 0.000017 & 34.8 & 24.4 & 23.1 & 24.7 & 22.8 & 173.5 \\ 
$f_{4}$ & 5.637247 $\pm$ 0.000019 & 31.3 & 22.3 & 20.9 & 22.3 & 20.6 & 156.8 \\ 
$f_{5}$ & 7.8995 $\pm$ 0.0004 & 1.4 & 1.1 & 1.1 & 0.9 & 0.9 & 7.0 \\ 
$f_{6}$ & 6.2434 $\pm$ 0.0003 & 1.4 & 1.0 & 0.9 & 1.2 & 1.3 & 8.4 \\ 
$f_{7}$ & 6.2618 $\pm$ 0.0002 & 2.8 & 2.0 & 1.8 & 2.4 & 1.9 & 15.2 \\ 
$f_{8}$ & 7.2026 $\pm$ 0.0004 & 1.4 & 1.3 & 1.2 & 1.0 & 0.7 & 7.3 \\ 
$f_{9}$ & 7.9131 $\pm$ 0.0004 & 1.3 & 1.0 & 0.8 & 1.3 & 0.9 & 6.9 \\ 
$f_{10}$ & 7.9307 $\pm$ 0.0004 & 1.3 & 1.3 & 1.2 & 0.8 & 1.0 & 7.0 \\ 
\\
$f_{A}$ & 0.4307 $\pm$ 0.0003 & 3.0 & 1.5 & 1.6 & 1.3 & 1.2 & 5.7 \\ 
$f_{C}$ & 0.3242 $\pm$ 0.0002 & 3.8 & 2.4 & 2.1 & 3.3 & 3.2 & 11.7 \\
$f_{D}$ & 0.6870 $\pm$ 0.0004 & 2.2 & 1.3 & 1.5 & 0.8 & 1.1 & 4.4 \\
$f_{E}$ & 0.2648 $\pm$ 0.0004 & 2.4 & 1.3 & 1.5 & 1.6 & 1.0 & 5.2 \\
$f_{F}$ & 0.2951 $\pm$ 0.0004 & 3.2 & 1.7 & 1.7 & 0.8 & 1.1 & 4.5 \\
$f_{G}$ & 0.3894 $\pm$ 0.0004 & 2.4 & 1.6 & 1.6 & 0.9 & 0.7 & 4.3 \\
$f_{H}$ & 0.3395 $\pm$ 0.0005 & 2.4 & 1.3 & 1.3 & 0.7 & 1.0 & 4.0 \\ 
\\
$f_1-f_2$ & 0.1093473 & 3.4 & 2.0 & 2.4 & 0.6 & 1.0 & 5.0\\
$f_1-f_3$ & 0.1432171 & 2.1 & 1.1 & 1.3 & 0.6 & 0.5 & 3.6\\
$f_{3}+f_4$ & 11.257294 & 0.7 & 0.8 & 0.5 & 0.6 & 0.3 & 3.6 \\ 
$f_{2}+f_3$ & 11.273963 & 2.3 & 1.5 & 1.7 & 1.9 & 1.6 & 13.0 \\ 
$f_{2}+f_4$ & 11.291164 & 0.6 & 0.6 & 0.5 & 0.5 & 0.5 & 3.5 \\
$f_{1}+f_3$ & 11.383311 & 9.3 & 6.9 & 6.6 & 7.1 & 6.5 & 49.2 \\ 
$f_{1}+f_4$ & 11.400511 & 9.7 & 7.1 & 6.6 & 6.8 & 6.3 & 47.1 \\ 
$f_{1}+f_2$ & 11.417181 & 12.1 & 9.1 & 8.5 & 8.6 & 8.2 & 58.7 \\ 
$2f_{1}$    & 11.526528 &  4.3 & 2.9 & 2.6 & 2.9 & 2.6 & 18.5 \\ \\

$f_1+f_{2}-f_3$ & 5.7971337 & 0.8 & 0.9 & 0.7 & 1.0 & 0.7 & 5.6 \\ 
$f_{2}+f_3+f_4$ & 16.911211 & 0.7 & 0.4 & 0.6 & 0.5 & 0.4 & 3.5 \\
$2f_{2}+f_3$    & 16.927880 & 1.4 & 1.1 & 1.0 & 0.5 & 0.4 & 4.4 \\
$f_{1}+2f_3$    & 17.003358 & 0.8 & 0.6 & 0.5 & 0.5 & 0.4 & 3.5 \\ 
$f_1+f_{3}+f_4$ & 17.020558 & 1.0 & 0.6 & 0.7 & 1.1 & 0.5 & 4.9 \\
$f_{1}+f_2+f_3$ & 17.037227 & 3.7 & 2.6 & 2.2 & 2.3 & 2.1 & 19.5 \\ 
$f_{1}+f_2+f_4$ & 17.054428 & 1.0 & 0.9 & 1.0 & 1.3 & 0.7 & 8.3 \\ 
$f_{1}+2f_2$ & 17.071097 & 0.8 & 0.6 & 0.7 & 0.2 & 0.6 & 3.9 \\ 
$2f_{1}+f_3$ & 17.146575 & 1.2 & 0.8 & 0.9 & 1.1 & 0.8 & 7.4 \\ 
$2f_{1}+f_4$ & 17.163775 & 1.5 & 1.3 & 0.9 & 0.9 & 0.9 & 9.1 \\ 
$2f_1+f_{2}$ & 17.180444 & 1.3 & 1.2 & 0.8 & 1.5 & 1.1 & 8.5 \\ 
\\
$f_1+2f_{2}+f_3$ & 22.691144 & 0.4 & 0.4 & 0.3 & 0.6 & 0.5 & 3.9 \\ 
$2f_{1}+f_2+f_3$ & 22.801022 & 1.6 & 0.9 & 1.1 & 1.1 & 1.1 & 10.3 \\ 
$2f_{1}+2f_2$ & 22.834361 & 0.3 & 0.2 & 0.2 & 0.6 & 0.2 & 4.0 \\ 
\hline
\end{tabular}
\end{center}
\label{tab:freq}
\end{table*}

The residuals from this solution were searched for other signals that 
may be intrinsic to the star. We analysed the amplitude-scaled residuals 
in the different filters (whereby the $u$ data were divided by 1.5, the 
$v$ and $B$ data divided by 1.05, and the $R$ data multiplied by 1.05 to 
scale them to amplitudes similar to that in the $y$ filter), and using 
different combinations of data sets. Even though some interesting 
possibilities for further intrinsic pulsation frequencies offered 
themselves, we prefer to err on the side of caution and stop the 
frequency search at this point.

Nevertheless, it is interesting to note that the residual scatter in the 
light curves (6.4 mmag per point in $u$, 5.2 mmag/point in $v$, 4.5 
mmag/point in $y$, 5.5 mmag/orbit in $B$, and 4.2 mmag/orbit in $R$) is 
larger than the precision of the observations. This may be partly 
explicable by data reduction imperfections, but residual intrinsic 
stellar variability is likely also involved.

\section{Mode identification}

BRITE-Constellation was designed to obtain multicolour time-series 
photometry specifically to provide identifications of the spherical 
degree $l$ of the modes of pulsating stars (Weiss et al. 2014). 
Daszy{\'n}ska-Daszkiewicz (2008) provided diagnostic diagrams to perform 
such mode identifications from a comparison of the amplitude ratios and 
phase shifts between the two BRITE passbands.

\begin{figure}
\includegraphics[width=85mm,viewport=-00 15 270 600]{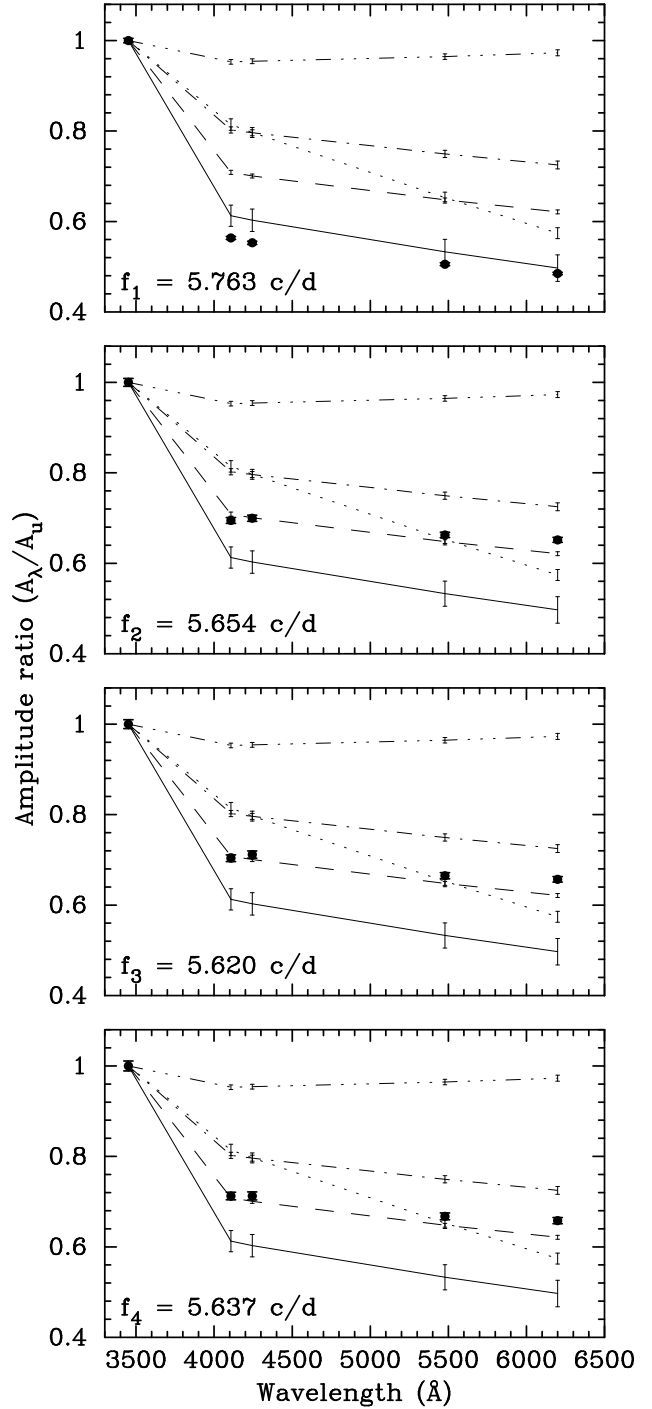}
\caption[]{Identification of the spherical degree of the four strongest 
pulsation modes of $\nu$~Eri. The filled circles with error bars are the 
observed amplitude ratios, in order of increasing wavelength $u/u$, $v/u$, 
$B/u$, $y/u$ and $R/u$. The full lines are theoretical predictions for 
radial modes, the dashed lines for dipole modes, the dashed-dotted lines 
for quadrupole modes, the dotted lines for $\ell=3$ modes, and the 
dashed-dot-dot-dotted lines are for $\ell=4$. The thin error bars 
delineate the uncertainties in the theoretical amplitude ratios.}
\label{fig:mid}
\end{figure}

The four strongest modes of $\nu$~Eri are both spectroscopically and 
photometrically uniquely identified (Handler et al. 2004, De Ridder et 
al. 2004, Daszy{\'n}ska-Daszkiewicz \& Walczak 2010): the strongest mode 
is radial (spherical degree $l=0$) and the high-amplitude triplet 
consists of dipole modes ($l=1$). In Fig.~\ref{fig:mid}, the observed 
$uvyBR$ amplitude ratios are compared to theoretical predictions for the 
used Str\"omgren and BRITE passbands for these four modes.

The theoretical amplitude ratios were computed with the method of Balona 
\& Evers (1999), using the nonadiabatic $f$ parameter (which is the 
ratio of the amplitude of the bolometric flux variations to the radial 
displacement at the photosphere level) calculated from theoretical 
models, and an arbitrary user supplied value of the pulsational radial 
velocity variation to be translated to photometric amplitudes. Pulsation 
modes with nonadiabatic frequencies between $5.3-6.0$\,\cd were 
considered because the amplitude ratios are sensitive to the radial 
overtone of the modes. This approach is in principle the same as that of 
Daszy{\'n}ska-Daszkiewicz (2008), and we have verified that the outcome 
of our calculations (based on older model atmospheres) is consistent 
with hers. The parameters used for $\nu$~Eri are consistent with the 
seismic models by Pamyatnykh et al. (2004), i.e. the $f$ parameters were 
taken from stellar models of 9.5 and 10 M$_{\sun}$ in a temperature 
range of $22000 \pm 600$~K. The results from non-LTE model atmosphere 
analysis of $\nu$~Eri by Nieva \& Przybilla (2012, $T_{\rm eff} = 22000 
\pm 400$~K, log $g=3.85\pm0.05$) are in excellent agreement with this 
choice. The ranges in the adopted parameters translate into the error 
bars of the theoretical amplitude ratios indicated in 
Fig.~\ref{fig:mid}.

\begin{figure*}
\includegraphics[width=80mm,angle=270,viewport=7 06 277 576]{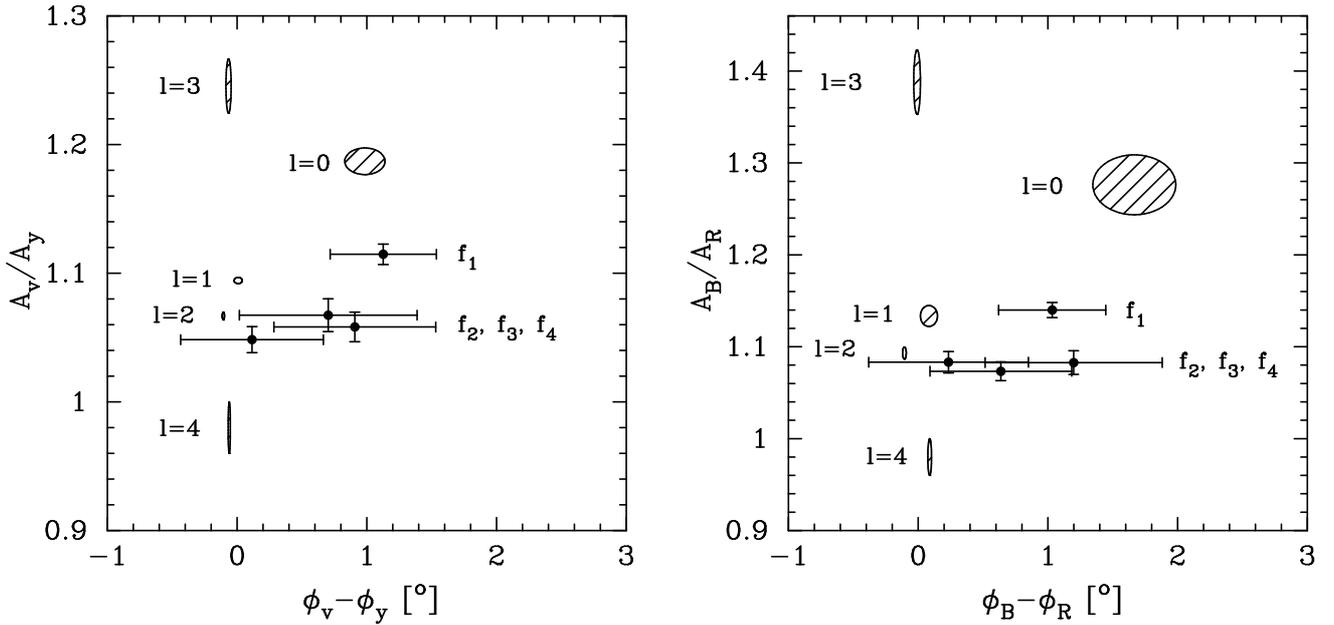}
\caption[]{Identification of the spherical degree of the four strongest 
pulsation modes of $\nu$~Eri from optical filters only. The shaded areas 
delineate the theoretically predicted domains of modes depending on 
their spherical degree (note that some of these are very small), whereas 
the filled circle with the error bars are the observed loci of the 
radial mode $f_1$ and the dipole modes $f_2 - f_4$.}
\label{fig:2cmid}
\end{figure*}

There are no big surprises in the mode identification: the literature 
results are confirmed. However, a closer look at the visual amplitude 
ratios alone (i.e. excluding the $u$ filter) reveals that for both $l=0$ 
and $l=1$ models predict a steeper increase in amplitude towards shorter 
wavelengths than observed. This is better seen in a comparison of 
observed and theoretical amplitude ratios and phase shifts for the 
optical $v/y$ and $B/R$ filter combinations (Fig.~\ref{fig:2cmid}) 
computed with the same parameters as specified above; we shall discuss 
this result in Sect. 5.

Because of their low amplitudes, identifying the $l$ value of the 
remaining modes from our data does not result in an improvement over 
results in the literature. We refer to the study by 
Daszy{\'n}ska-Daszkiewicz \& Walczak (2010) for observational 
identifications of these modes.


\section{Discussion}

Compared to previous results, in particular to the work by JHS, there 
are several differences. These authors reported 36 frequencies in their 
light curves, of which two originated from the main comparison star. The 
present study revealed 40 frequencies of light variation, all of them 
due to $\nu$~Eri.

Among the independent frequencies that would correspond to 
$\beta$~Cep-type p modes, the present work has ten in common with JHS. 
We do not detect their two weakest signals in this domain, at 6.2236 and 
6.7332\,\cd. The first signal has been found in other studies, 
particularly in the spectroscopic analysis by Aerts et al. (2004), which 
raises the suspicion that its amplitude has dropped below the present 
detection level. Searching for a signal at this frequency yielded a null 
result. Concerning the 6.7332\,\cd signal, we cannot find a trace of it.

As just mentioned, some of the stellar pulsational signals change their 
amplitudes, apparently even to the extent that they can sometimes be 
detected, but sometimes disappear beyond the observational detection 
threshold. The differences between the independent pulsation frequencies 
and amplitudes as derived by JHS and us, and common to both studies, are 
summarized in Table~\ref{tab:frevar}.

It becomes clear that the amplitude of the strongest mode has 
considerably dropped compared to the results by JHS - by some 6\%
over the last decade. Therefore these amplitude 
variations are not a simple consequence of the interference of 
low-amplitude pulsation modes with observational noise, but are 
intrinsic to the star. On the other hand, the amplitudes of the $f_2 - 
f_4$ triplet appeared fairly constant, aside from a slight increase of 
the amplitude of mode $f_2$ compared to the study by JHS. 

We recall that Handler et al. (2004) already argued in favour of some 
amplitude variation occurring in the pulsation spectrum of $\nu$~Eri. 
In this context it is interesting to note that an amplitude {\it 
increase} of $f_1$ with respect to the measurements by van Hoof (1961), 
analysed by Cuypers \& Goossens (1981), had occurred, and that the 
relative amplitudes within the $f_2 - f_4$ triplet had also changed.

As regards to the remaining modes, decreases in the amplitudes of modes 
$f_5$, $f_6$ and $f_A$ are also present at a significant level. The 
amplitudes of modes $f_7$ through $f_{10}$ stayed the same within the 
errors, keeping in mind that our error bars likely understimate the real 
uncertainties.

\begin{table}
\caption[]{Changes in the pulsational amplitudes and frequencies of
$\nu$~Eri between the study of JHS and the present one. The values were
calculated as JHS {\it minus} present.}
\begin{center}
\scriptsize
\begin{tabular}{lcccc}
\hline
ID & $\Delta$ f & $\Delta$$u$  &$\Delta$$v$  &$\Delta$$y$  \\
 & ($10^{-3}$\cd) & (mmag) & (mmag) & (mmag) \\
\hline
$f_{1}$ & $-$0.019 $\pm$ 0.012 & $-$4.8 $\pm$ 0.3 & $-$2.5 $\pm$ 0.2 & $-$2.3 $\pm$ 0.2\\
$f_{2}$ & +0.040 $\pm$ 0.015 & +1.1 $\pm$ 0.3 & +0.4 $\pm$ 0.2 & +0.8 $\pm$ 0.2\\
$f_{3}$ & +0.028 $\pm$ 0.017 & $-$0.2 $\pm$ 0.3 & $-$0.1 $\pm$ 0.2 & $-$0.1 $\pm$ 0.2\\
$f_{4}$ & +0.000 $\pm$ 0.019 & $-$0.5 $\pm$ 0.3 & +0.0 $\pm$ 0.2 & $-$0.1 $\pm$ 0.2\\
$f_{5}$ & +1.3 $\pm$ 0.4 & $-$2.2 $\pm$ 0.3 & $-$1.5 $\pm$ 0.2 & $-$1.4 $\pm$ 0.2\\
$f_{6}$ & $-$0.4 $\pm$ 0.3 & $-$1.6 $\pm$ 0.3 & $-$0.9 $\pm$ 0.2 & $-$1.2 $\pm$ 0.2\\
$f_{7}$ & $-$1.1 $\pm$ 0.2 & +0.0 $\pm$ 0.3 & +0.0 $\pm$ 0.2 & +0.0 $\pm$ 0.2\\
$f_{8}$ & +1.7 $\pm$ 0.4 & +0.0 $\pm$ 0.3 & +0.4 $\pm$ 0.2 & +0.3 $\pm$ 0.2\\
$f_{9}$ & $-$0.7 $\pm$ 0.4 & $-$0.4 $\pm$ 0.3 & $-$0.1 $\pm$ 0.2 & $-$0.4 $\pm$ 0.2\\
$f_{10}$ & +0.8 $\pm$ 0.4 & +0.1 $\pm$ 0.3 & +0.4 $\pm$ 0.2 & +0.3 $\pm$ 0.2\\
$f_{A}$ & $-$2.1 $\pm$ 0.3 & $-$1.1 $\pm$ 0.3 & $-$1.0 $\pm$ 0.2 & $-$0.9 $\pm$ 0.2\\
\hline
\end{tabular}
\normalsize
\end{center}
\label{tab:frevar}
\end{table}

Turning to the independent low frequencies ($\nu$~Eri is a hybrid p and 
g-mode pulsator), considerable changes have occurred since the work of 
JHS. A previously unknown signal $f_C$ now dominates the low-frequency 
domain. On the other hand, the low-frequency signal $f_B$ by JHS is no 
longer detectable at all, and the amplitude of signal $f_A$ has dropped 
to about 2/3 of its previous value. The frequency of this signal has 
also considerably shifted. The biggest change compared to the results of 
JHS, however, is the detection of a total of six new low-frequency 
independent modes ($f_C - f_H$). Had these signals been present with 
similar amplitudes in the old data, most of them would have been 
detected, as a re-examination of this data set shows. One may 
suspect that the appearance of these g modes may be related to the 
decrease in amplitude of several of the p modes, in the sense that the 
pulsational energy lost by the p modes became available to raise the 
amplitudes of these g modes.

The newly detected g modes make it tempting to look for equal frequency 
splittings indicative of the rotation rate in the stellar interior, 
particular when recalling the result by Pamyatnykh et al. (2004) that 
$\nu$~Eri rotates some four times more rapidly near the core than on its 
surface. However, the theoretically predicted g-mode frequency spectrum 
(e.g., see Fig. 6 of Daszy{\'n}ska-Daszkiewicz \& Walczak 2010) of $\nu$ 
Eri is very dense. Even in the small frequency region between 0.26 -- 
0.43\,\cd in which we detected six g modes, models predict 13 radial 
overtones of $l=1$ g modes. Any claim of rotational splitting in the 
g-mode domain must therefore remain in the realm of speculation. On the 
other hand, it can at the same time also be speculated that the apparent 
change in frequency of mode $f_A$ may actually be a manifestation of a 
different mode present in the current data set.

Moving on to the combination frequencies, our results are mostly 
consistent with those by JHS (apart from some amplitude variations) as 
far as the first-order combinations are concerned, apart from us not 
detecting $2f_2$, $f_1+f_5$ and $f_1-f_4$, but $f_1-f_3$ instead. We 
detect all the higher-order combination frequencies as did JHS, and 
add five more due to our lower detection limit.

Two $\delta$ Scuti-type oscillation frequencies (10.874 and 17.252\,\cd) 
originating from the comparison star $\xi$~Eri were reported by JHS; the 
lower-frequency of them was preliminarily mentioned by Handler et al. 
(2004). The latter signal can be discerned in our present ground-based 
photometry using the same comparison star, but does not occur at a 
significant level. We were unable to find the higher-frequency variation 
reported by JHS. We caution that our new ground-based data are 
considerably fewer than those reported by JHS, which may explain these 
differences.

We reported that the theoretically predicted pulsation 
amplitude-wavelength relation in the visual range is steeper than that 
observed (Fig.~\ref{fig:mid}). This raises the suspicion that the filter 
bandpasses we used for the calculations may be inaccurate. That the 
effect is systematic over the visual wavelength range covered by our 
measurements (4100 $-$ 6200\,\AA), i.e. not only one or two filters in 
the optical are affected, argues against such an interpretation, 
however. The same argument refutes that the problem stems from 
imperfections in the data reduction procedure; furthermore the 
ground-based and space photometry were reduced independently.

Figure~\ref{fig:2cmid} shows the discrepancy more drastically. The 
comparison of the theoretical loci of low-$l$ pulsation modes and the 
observed ones using combinations of two optical filters only and the 
observed positions of the well-identified modes $f_1 - f_4$ shows some 
large discrepancies as far as the amplitude ratios are concerned. Using 
these diagrams, the radial mode $f_1$ would rather be identified as a 
dipole, whereas the $l=1$ modes $f_2 - f_4$ are located where $l=2$ 
modes would be expected. Since this problem does arguably not originate 
from poorly known filter bandpasses, another explanation in physical 
terms needs to be sought.

For instance, rapid stellar rotation would change the temperature and 
mode amplitude distribution on the stellar surface. Hence amplitude 
ratios between different filters would become dependent on the 
inclination of the rotation axis to the line of sight, the rotation 
rate, and the azimuthal order of the modes and can deviate considerably 
from the non-rotating case (e.g., Reese et al. 2013). However, $\nu$~Eri 
is an intrinsically slow rotator (Pamyatnykh et al. 2004) with a surface 
rotation period of about two months, and such a possibility can hence be 
ruled out.

The pulsation amplitudes of $\nu$~Eri are relatively high and therefore 
harmonic and combination frequencies are observed. One may therefore 
surmise that nonlinear effects could modify the amplitudes, which would 
be unaccounted for by the models used to compute the theoretical 
amplitude ratios. Indeed, the dominant modes of another $\beta$~Cep star 
with prominent harmonics and combination frequencies, 12~Lac (Handler et 
al. 2006) also have smaller amplitude ratios in the optical than 
theoretically predicted. On the other hand, stars with weak or no 
harmonics and combination frequencies with well-identified oscillations 
(e.g, IL Vel and KZ Mus, Handler et al. 2003., V2052 Oph, Handler et al. 
2012) do not seem to show this effect, whereas it may be suspected in 
others (V836 Cen, Dupret et al. 2004).

Another possibility is that there is a problem with the theoretical 
predictions of the pulsation amplitude vs. wavelength relation. We may 
for instance just have used non-optimal stellar input parameters. 
However, in such a situation it would be hard to understand why the 
amplitude ratios with respect to the Str\"omgren $u$ filter were quite 
well reproduced. For instance, an incorrectly adopted effective 
temperature would cause a systematic change of the predicted amplitude 
ratios within all passbands as they are all located in the 
Rayleigh-Jeans tail of the stellar spectral energy distribution. We 
recall that the effective temperature adopted for our calculations is in 
perfect agreement with recent and modern spectral analyses (Nieva \& 
Przybilla 2012).

One more possible culprit would be a poor choice of the complex $f$ 
parameter that is a vital ingredient for the computation of theoretical 
pulsation amplitude ratios. As pointed out by, e.g., 
Daszy{\'n}ska-Daszkiewicz \& Walczak (2010), this parameter is very 
sensitive to the metal content and opacities in the stellar atmosphere. 
We have therefore made some experiments with modifying the $f$ 
parameter, e.g. by adopting empirically determined values 
(Daszy{\'n}ska-Daszkiewicz \& Walczak 2010) instead of the ones 
predicted by stellar models. Our experiments did not bear fruit, but 
more trials in this direction are advised. In any case, any seismic 
model that contains or proposes modifications to the $f$ parameter 
should also reproduce the observed amplitude ratios.

To conclude, we cannot offer a good explanation for the discrepancy 
between the observed and theoretical pulsational amplitude ratios for 
the dominant modes of $\nu$ Eri in the optical at this point. However, 
we do want to stress that identifying pulsational modes of $\beta$ Cep 
stars from optical amplitude ratios (and phase shifts) alone may be 
misleading, at least until this problem has been resolved.

\section{Summary and conclusions}

We have reported the first combined analysis of simultaneous 
BRITE-Constellation and ground-based photometry. The measurements were 
acquired for the well-studied $\beta$~Cep star $\nu$~Eri, with a 
combined duration of 173.5~d. This exemplary data set was used to show 
the power of joining these two types of observations, because they 
complement each other well in terms of suppressing aliasing problems in 
the data; aliasing is practically non-existent if the data can be 
carefully analysed together.

Performing such a frequency analysis, we detected 40 signals in the 
light curves. Among these are six newly discovered gravity-mode 
frequencies. Their detection is a consequence of the combination of, 
firstly, the lower noise level in the BRITE data, which amounts to 72\% 
between $0-1$\,\cd and 80\% between $5-9$\,\cd of that in the combined 
$2002 - 2004$ ground-based multisite photometry (JHS). The second factor 
is the evolution of the stellar pulsation amplitudes. Such changes in 
the stellar mode spectrum make frequency analyses more challenging, but 
are also an asset as repeated observations may reveal more seismic 
information. Unfortunately, the data set reported here cannot be 
analysed together with those from the multisite campaigns carried out 
more than a decade ago (JHS) as the temporal gap between the two 
observational campaigns is too large, and just because the amplitude 
variations have occurred. However, at least some of the BRITE satellites 
will continue to observe $\nu$~Eri, and eventually this gap can be 
bridged.

We showed that the observed pulsational amplitude ratios for the 
strongest modes are consistent with their previous identifications. 
However, the amplitude-wavelength relation in visual passbands is 
flatter than that predicted from theory. Although we cannot offer a good 
explanation for this discrepancy at this point, we suggest that mode 
identification of $\beta$~Cep stars should not be done in the optical 
alone; a UV band or radial velocities are necessary.

The pulsation amplitude-wavelength relation observed for $\nu$~Eri 
requires theoretical explanation to derive reliable mode identifications 
from BRITE data alone. Furthermore, the additional gravity modes 
detected in the present work call for explanation, as current pulsation 
models have difficulties to excite them. An upcoming theoretical study 
by Daszy{\'n}ska-Daszkiewicz et al. (2016) will address these 
questions.

\section*{ACKNOWLEDGEMENTS}

This work has been supported by the Polish NCN grants 
2011/01/B/ST9/05448, 2011/01/M/ST9/05914, 2011/03/B/ST9/02667 and 
2015/18/A/ST9/00578. TR acknowledges support from the Canadian Space 
Agency grant FAST. GAW and SMR acknowledge Discovery Grant support from 
the Natural Science and Engineering Research Council (NSERC) of Canada. 
AFJM is grateful for financial aid from NSERC and FQRNT (Quebec). KZ 
acknowledges support by the Austrian Fonds zur F\"orderung der 
wissenschaftlichen Forschung (FWF, project V431-NBL). BTr operations are 
supported through a Canadian Space Agency (CSA) Academic Development 
grant.

\bsp

\end{document}